\documentclass{actasen}
\begin{document}

\setcounter{page}{1}
\Volume{2015}{}
\runheading{Hiroaki Abuki}%

\title{Inhomogeneous chiral phases\\
in two-flavor quark matter}
\footnote{\noindent\hspace*{5mm}$^{\bigtriangleup}$ abuki@auecc.aichi-edu.ac.jp}
\enauthor{Hiroaki Abuki$^{\,1,\,2,\,\bigtriangleup}$}{%
$^{1}$Faculty of Education, Aichi University of Education, \\
Hirosawa 1, Igaya-cho, Kariya 448-8542, Japan\\
$^{2}$Research and Education Center for Natural Sciences, Keio University, \\
Hiyoshi 4-1-1, Yokohama, Kanagawa 223-8521, Japan}

\abstract{
We present a systematic study of the phase structure of QCD in a
generalized Ginzburg-Landau framework.
We find, going up in density, a strongly interacting matter might go
through the ``pion crystal'', an exotic inhomogeneous chiral phase
before reaching the full restoration of symmetry.}

 \maketitle

The possibility of spatially inhomogeneous realizations of chiral
symmetry breaking in quark matter has recently been the subject of
extensive research \cite{1}.
We here report our recent study \cite{2,3} on the inhomogeneous chiral
phases near the (tri-)critical point, (T)CP hereafter.
A particular focus is put on the impact of isospin asymmetry since
up and down quarks are not equally populated in a realistic situation
such as expected in the interior of compact stars.
The method we use is the generalized Ginzburg-Landau (gGL) approach.
We show that in some region of phase diagram, quark matter may be
realized in a form of solitonic charged pion crystal (SPC).

The gGL potential at the minimal order describing the (T)CP at
$\mu_\mathrm{I}=0$ is \cite{2}
\begin{equation}
\begin{array}{rcl}
 \omega({\bf x})&=&\displaystyle-h\sigma+\frac{\alpha_2}{2}\phi^2%
 +\frac{\alpha_4}{4}(\phi^4+(\nabla\phi)^2)\\[2ex]
 &&\displaystyle%
 +\frac{\alpha_6}{6}\left(\phi^6+3[\phi^2(\nabla\phi)^2%
-(\phi\cdot\nabla\phi)^2]%
 +5(\phi\cdot\nabla\phi)^2+\frac{1}{2}(\Delta\phi)^2\right),
\end{array}
\label{eq:gGL}
\end{equation}
where we utilized the chiral four-vector notation
$\phi({\bf x})=(\sigma({\bf x}),\pi_1({\bf x}),\pi_2({\bf
x}),\pi_3({\bf x}))$ with $\sigma({\bf x})\sim\langle\bar{q}q\rangle$
the scalar condensate and ${\pi}_a({\bf
x})\sim\langle\bar{q}i\gamma_5\tau_a q\rangle$ ($a=1,\,2,\,3$) the
pseudoscalar condensates. 
$h$ is an external field due to a small current quark mass, by which 
the chiral O(4) symmetry is broken to its diagonal subgroup, the
isospin O(3).
The effect of isospin asymmetry can be accommodated via introducing 
the isospin chemical potential $\mu_{\mathrm{I}}$.
It further breaks O(3) to U(1) symmetry, the rotation about the isospin
third axis.
The terms that should added to are summarized as
\begin{equation}
\delta\omega({\bf x})=\frac{\beta_2}{2}\bm{\pi}_\perp^2%
+\frac{\beta_4}{4}\bm{\pi}_\perp^4+\frac{\beta_{4b}}{4}(%
\phi^2-\bm{\pi}_\perp^2)\bm{\pi}_\perp^2+\frac{\beta_{4c}}{4}%
(\nabla\bm{\pi}_\perp)^2,
\label{eq:deltaomega}
\end{equation}
where we have defined the charged pion doublet
$\bm{\pi}_\perp=(\pi_1,\pi_2)$.
If the system is in the charged pion condensate (PC) with
$|\bm{\pi}_\perp|\ne0$,
that is when $\mathrm{U}(1)$ symmetry is spontaneously broken.
Within the expansion in $\mu_\mathrm{I}$,
we find $\beta_2=-\frac{1}{2}\mu_\mathrm{I}^2\alpha_4$,
$\beta_{4}=-2\mu_\mathrm{I}^2\alpha_6$, $\beta_{4b}=-2\mu_\mathrm{I}^2%
  \alpha_6$, and $\beta_{4c}=-\frac{4}{3}\mu_\mathrm{I}^2\alpha_6$.
The gGL potential $\omega({\bf x})+\delta\omega({\bf x})$ is 
characterized by five parameters, $\alpha_2$, $\alpha_4$, $\alpha_6$,
$h$, and $\mu_\mathrm{I}^2$.
Assuming $\alpha_6>0$ for the stability we can replace $\alpha_6$ with $1$
adopting the convention that every quantity
with an energy dimension is to be measured in the unit
$\alpha_6^{-1/2}$.
Then via scaling $\mu_\mathrm{I}\to h^{1/5}\mu_\mathrm{I}$, $\phi\to
h^{1/5}\phi$, ${\bf x}\to{\bf x}^{-1/5}$,
$\alpha_2\to\alpha_2 h^{4/5}$, $\alpha_4\to\alpha_4 h^{2/5}$, 
we can get rid of $h$ in $\omega$ apart from an overall factor $h^{6/5}$.
Accordingly we are left with three parameters $\alpha_2$, $\alpha_4$ and
$\mu_\mathrm{I}^2$ which are to be measured in units $h^{4/5}$,
$h^{2/5}$ and $h^{2/5}$ respectively.

\begin{figure}[tbp]
\centering
%\sidecaption
\includegraphics[width=0.65\textwidth,clip]{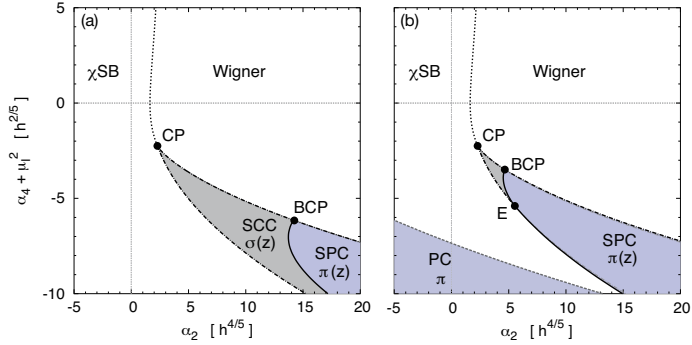}
\caption{The gGL phase diagram for nonvanishing isospin densities:
(a)~for $\mu_\mathrm{I}^2=0.01$, and (b)~for $\mu_{\mathrm{I}}^2=0.1$.
}
\label{fig-2}
\end{figure}

Now we can evaluate the phase structure in the
$(\alpha_2,\alpha_4)$-plane. 
The result is displayed in Fig.~\ref{fig-2};
(a) is for $\mu_\mathrm{I}^2=0.01$, and (b) is for $\mu_\mathrm{I}^2=0.1$.
The former roughly corresponds to $\mu_\mathrm{I}\sim 50\,$MeV,
and the latter to $\mu_\mathrm{I}\sim 150\,$MeV \cite{3}.
The vertical axis is shifted by $\mu_\mathrm{I}^2$ so as to make
the trivial shift of location of CP invisible.
The shift corresponds to higher $\mu$ and lower $T$ \cite{3}.
The chiral symmetry is broken in the $\chi$SB phase while it is nearly
restored in the ``Wigner'' phase; there is no phase boundary between
them for $\alpha_4$ larger than the value at the CP.
The phase labeled by ``SCC'' is the solitonic chiral condensate 
characterized by three parameters $b$, $k$ and $\nu$ as:
$
 \sigma(z)=k\nu^2\mathrm{sn}(b,\nu)\mathrm{sn}(kz-b/2,\nu)%
 \mathrm{sn}(kz+b/2,\nu)%
 +k\frac{\mathrm{cn}(b,\nu)\mathrm{dn}(b,\nu)}{\mathrm{sn}(b,\nu)}.
$
It provides a one-parameter family of solutions to the
equation $\frac{\delta\Omega}{\delta\sigma(z)}=0$ \cite{3}.
``SPC'' is the solitonic charged pion crystal condensate, a charged pion
analog to the SCC state.
It is defined by $\sigma\ne0$ and $\pi_1(z)=k\nu\mathrm{sn}(kz,\nu)$.
In this phase, the charged pion component is oscillating in the
homogeneous sea of scalar condensate.
The SPC and Wigner phases are separated by a continuous
second-order phase transition, 
At the phase boundary, a charged pion mode at a finite wavevector
becomes unstable; the amplitude of the mode grows exponentially with
time.
Nonvanishing $\mu_\mathrm{I}^2$ is responsible for this because it gives a
negative contribution to the gradient term $(\nabla\bm{\pi}_\perp)^2$ for
$\beta_{4c}=-\frac{4}{3}\mu_\mathrm{I}^2\alpha_6<0$.
By contrast, the phase transition between the SPC and SCC phases is
first-order.
As a consequence, there is a bicritical point denoted by ``BCP'' where
two second-order phase transitions and a first-order phase transition
meet up.
A notable change in the topology of phase diagram (b) from (a) is
the appearance of continent of PC in the depths of homogeneous
$\chi$SB phase.
This is triggered by the cross-term $\alpha_4\sigma^2%
\bm{\pi}_\perp^2$ in the gGL potential Eq.~(\ref{eq:deltaomega}) 
which drives the instability in the charged pion mode when
$\alpha_4<0$ and $\sigma^2$ is large.

\begin{figure}[tbp]
\centering
\includegraphics[width=0.7\textwidth,clip]{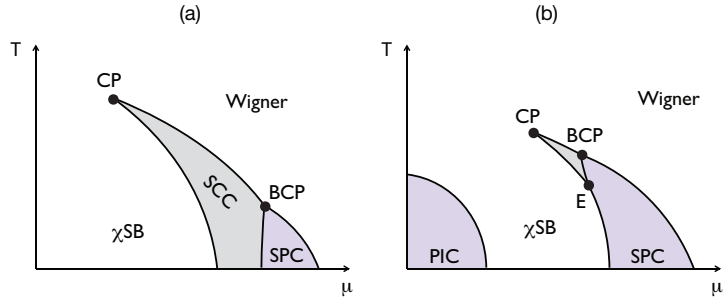}
\caption{Schematic picture of QCD phase diagram for $\mu_\mathrm{I}\ne
 0$: (a)~for small $\mu_\mathrm{I}^2$ which shares the same topology
 with Fig.~\ref{fig-2}(a), and (b)~for large $\mu_I^2>m_{\pi}^2$, where
 the situation corresponds to Fig.~\ref{fig-2}(b).
}
\label{fig-3}
\end{figure}

Let us finally draw a schematic QCD phase diagram in
$(\mu,T)$-plane, though it stays at the conceptual level.
This is done in Fig.~\ref{fig-3} where (a) and (b) are 
mapped into from those in Fig.~\ref{fig-2}.
We utilized following two guidelines: 
(1) The topology in gGL phase diagram in $(\alpha_2,\alpha_4)$-plane
should be kept. 
(2) In the vicinity of CP, the $\alpha_4$ ($\alpha_2$) axis points to
higher $T$ and lower (higher) $\mu$ direction \cite{2}.
In both cases (a) and (b), going up in density the system goes
through the SPC phase before realizing an entire restoration of chiral
symmetry.
CP moves towards higher chemical potential and lower temperature with
increasing $\mu_\mathrm{I}$.
As soon as $\mu_\mathrm{I}$ surpasses the mass of charged pion, the PC would be
realized even in the absence of net quarks.
For this reason the PC continent in $(\alpha_2,\alpha_4)$-plane is
mapped on to the area including the origin.

\vspace*{1ex}
\noindent
{\em Acknowledgements.} The author would like to express his sincere thanks to
all the organizers of QCS2014 for their warm hospitality during the workshop.

\end{document}